\documentclass[aps,prb,twocolumn,amsmath,showkeys,amssymb,superscriptaddress]{revtex4-2}
\usepackage{xcolor}

\usepackage{amscd}
\usepackage{graphicx}
\usepackage{dcolumn}
\usepackage{bm}

\newcommand{\bd}[1]{ \mbox{\boldmath $#1$}}

\begin{document}


\title{Semiclassical study of single-molecule magnets and their quantum phase transitions}



\author{D. S. Lohr-Robles}
\affiliation{Instituto de Ciencias Nucleares, Universidad Nacional Aut\'onoma de M\'exico, A.P. 70-543, 04510 Mexico-City, Mexico}

\author{E. L\'opez-Moreno}
\affiliation{Facultad de Ciencias,  Universidad Nacional Aut\'onoma de M\'exico, 04510 Mexico-City, Mexico}

\author{P. O. Hess}
\affiliation{Instituto de Ciencias Nucleares, Universidad Nacional Aut\'onoma de M\'exico, A.P. 70-543, 04510 Mexico-City, Mexico}
\affiliation{Frankfurt Institute for Advanced Studies, J. W. von Goethe University, Hessen, Germany}


\date{\today}

\begin{abstract}
We present a study of systems of single-molecule magnets using a semiclassical analysis and catastrophe theory. Separatrices in parameter space are constructed which are useful to determine the structure of the Hamiltonians energy levels. In particular the Maxwell set separatrix determines the behavior of the ground state of the system. We consider an external magnetic field with two components, one parallel to the easy magnetization axis of the molecule and the other perpendicular to it. Using the fidelity and heat capacity we were able to detect the signals of the QPTs as a function of the magnetic field components.
\end{abstract}


\maketitle


\section{Introduction}\label{introduction}
The subject of single-molecular magnets (SMMs) is an area of growing experimental and theoretical research, ever since the first identification of the magnetic properties of $\mathrm{Mn_{12}}$-ac molecule at the start of the 1990s \cite{caneschi1991a,sessoli1993a}. The interest is due to the quantum effects exhibited in the small magnets \cite{gatteschi2003a}, which offer convincing possibilities for their use in quantum technologies \cite{leuenberger2001a,thiele2013a,mousolou2014a,burgess2015a,singh2016a,gaitaarino2019a,zabalalekuona2021a,morenopineda2021a}. For this reason a lot of effort has been put in engineering new examples of SMMs that work at higher temperatures over long times, using a diversity of approaches and materials \cite{soler2004a,milios2007a,cirera2009a,pugh2015a,lunghi2017a,liu2017a,cini2018a,singha2021a}. It has been shown that in order to synthesize SMMs that exhibit magnetization at high temperatures it must have a large ground state spin and uniaxial anisotropy \cite{ako2006a}. With this in mind a variety of experimental techniques are used in order to study the magnetic properties of the synthesized SMMs \cite{cornia2001a}, and new techniques are being proposed \cite{luneau2021a}.

The purpose of this paper is to contribute to the theoretical aspect of the study of SMMs, particularly in the effect of the anisotropy parameters and a varying external magnetic field to the dynamics of the SMMs. This is done performing a semiclassical analysis on the Hamiltonian of the system, with terms up to fourth order, by using coherent states \cite{hecht} as a trial function to calculate its expectation value. Then, employing methods of catastrophe theory \cite{gilmore}, we obtain parameter relations that delimit the qualitative behavior of the quantum energy levels of the Hamiltonian. The methods of catastrophe theory in similar models has already been proven useful in the area of nuclear physics in the interacting boson model (IBM) \cite{lopezmoreno1996a,garciaramos2014a,leviatan2020a}, in the semimicroscopical algebraic cluster model (SACM) \cite{lohrrobles2019a,lohrrobles2021a,lohrrobles2023a}, and in an effective QCD model \cite{lohrrobles2021a}. They have also been applied in a Lipkin-Glick-Meshkov model (LMG) and spin Hamiltonians \cite{castanos2006a,lopezmoreno2011a,lopezmoreno2014a,garciaramos2016a}, which are very related to the results presented here. In these models, catastrophe theory was successfully applied to describe quantum phase transitions (QPTs) \cite{sachdev} of first and second order as one of the parameters that describe the system is varied. The aim of the present research is to contribute with a framework for the study of QPTs in SMMs, which have been studied in a $\mathrm{Fe_8}$ magnet \cite{burzuri2011a}. The advantages of catastrophe theory are many, as it provides a systematic program to determine the dependence of the critical points of a function to the parameters that define it, while the information obtained is directly translated to quantum mechanical results. By the end of the analysis we are able to construct separatrices in parameter space for examples of SMMs with different anisotropy parameters and determine how a varying external magnetic field influence their dynamics, in a way that QPTs in such systems may be properly predicted and detected in physical variables.

In Section \ref{section2} we present the Hamiltonian that describes a SMM system with terms up to fourth order and begin the semiclassical analysis. The main concepts of catastrophe theory are also introduced. In Section \ref{section3} the results of the previous section are applied to particular examples of SMMs and their respective parameters spaces are constructed. We also present a schematic study of the effect of the fourth order parameters of the Hamiltonian to the parameter space. In Section \ref{section5} the ground state fidelity is used to map the region in parameter space where the QPTs occur. In Section \ref{section6} finite temperature phase transitions and their effect in the heat capacity of a SMM are studied. In Section \ref{conclusions} conclusions are discussed.

\section{Hamiltonian}\label{section2}
An effective model, at low temperature, consists in treating the whole system to be described by a single large spin, depending on the sum of ions in the center \cite{wilson2007a}. The spin Hamiltonian is composed of the components of the spin operator ${\bd S}=({\bd S}_x,{\bd S}_y,{\bd S}_z)$. The SMM has an axis of anisotropy, which corresponds to the axis favoured when an external magnetic field is applied. This is called the easy axis, and the $z$-axis is set parallel to it \cite{molecular}. In the plane perpendicular to it are the hard and intermediate axes, which correspond to the $x$ and $y$ axes, respectively. The Hamiltonian consists of a term describing the Zeeman interaction of the molecule with an external magnetic field ${\bd B}=(B_x,B_y,B_z)$. Second order terms are related to the composition of the molecule, for axial and transverse anisotropy, and the parameters are determined by experiment. When the experimental data cannot be adequately fitted, fourth order terms are included to obtain a better fit \cite{gatteschi2003a,wilson2007a}. The fourth order terms that are considered depend on the symmetry of the molecule: tetragonal, trigonal and orthorhombic \cite{gatteschi2003a,cornia2001a}.

With this in mind we will consider the following Hamiltonian to describe a SMM:
\begin{equation}\label{1.1}
{\bd H} = D {\bd S}^2_z + E({\bd S}^2_{x}-{\bd S}^2_{y}) + g \mu_B {\bd S}\cdot {\bd B} + {\bd H}_4 ,
\end{equation}
where $D$ and $E$ are the axial and transverse anisotropy parameters of the molecule, respectively; $g$ is the Land\'e factor, which is set to $g=2$, and $\mu_B$ is the Bohr magneton. The operator ${\bd H}_4$ refers to the fourth order anisotropy contributions and is given by:
\begin{eqnarray}\label{1.2}
{\bd H}_4 &=& B_4^0 {\bd O}_4^0 + B_4^2 {\bd O}_4^2 + B_4^3 {\bd O}_4^3 + B_4^4 {\bd O}_4^4 \cr
&=& B_4^0 \left(35{\bd S}_z^4 + (25-30S(S+1)){\bd S}_z^2 +3S^2(S+1)^2 \right.\cr
&&\left.-6 S(S+1) \right) \cr
& & +\frac{B_4^2}{4} \left((7{\bd S}_z^2 - S(S+1)-5)({\bd S}_{+}^2 + {\bd S}_{-}^2)\right.\cr
&& \left. +({\bd S}_{+}^2 + {\bd S}_{-}^2)(7{\bd S}_z^2 - S(S+1)-5) \right) \cr
& &  +\frac{B_4^3}{4} \left( {\bd S}_z ({\bd S}_{+}^3 + {\bd S}_{-}^3)+({\bd S}_{+}^3 + {\bd S}_{-}^3){\bd S}_z \right)\cr
& &  + \frac{B_4^4}{2}({\bd S}_{+}^4 + {\bd S}_{-}^4),
\end{eqnarray}
where ${\bd O}_4^k$ are the Stevens operators and $B_4^k$ are the corresponding parameters \cite{gatteschi2003a,cornia2001a}. Depending on the symmetry of the material the set of non-zero parameters are known: $(B_4^0\neq 0, B_4^4 \neq 0)$ for tetragonal symmetry, $(B_4^0\neq 0, B_4^3 \neq 0)$ for trigonal symmetry and $(B_4^0\neq 0, B_4^2\neq 0, B_4^4 \neq 0)$ for orthorhombic symmetry \cite{gatteschi2003a,cornia2001a}.

\subsection{Semiclassical potential and catastrophe theory}
The present Hamiltonian depends on $9$ parameters, six of them related to the anisotropy of the molecule and its symmetries, and which are fitted to experimental data, and three parameters are the components of the external magnetic field. The question we would like to answer is how does the change of the parameters, and in particular of the magnetic field, relates to changes in the eigenvalues of the Hamiltonian. To answer this question, we will perform a semiclassical analysis on the system. The semiclassical potential is obtained as the expectation value of the Hamiltonian (\ref{1.1}) in the atomic coherent states basis \cite{arecchi1972a}, which is defined as
\begin{equation}\label{1.3}
|\mu \rangle = \frac{1}{(1+|\mu|^2)^S}e^{\mu^{*}{\bd S}_+}|S -S\rangle ,
\end{equation}
where $|S M\rangle$ are the eigenfunctions of the ${\bd S}_z$ operator, with $|S -S\rangle$ the ground state, and ${\bd S}_{\pm}={\bd S}_x \pm i {\bd S}_y$ are the ladder operators. The coherent states parameters $\mu = e^{-i\phi}\tan (\theta /2) = (x+iy)/(1+z)$ are defined as the coordinates on a Bloch sphere \cite{arecchi1972a}. It is then possible to work with the semiclassical potential in angular coordinates or in cartesian coordinates.

The extrema of the semiclassical potential dictates the overall structure of the eigenvalues of the Hamiltonian (\ref{1.1}) as the magnetic field is varied. Catastrophe theory \cite{gilmore} provides us with a useful framework to determine the dependence of the critical points of the potential on the parameters. The bifurcation set and Maxwell set separatrices are constructed in parameter space delimiting regions of similar qualitative behavior for the potential. The \textit{bifurcation set} is the subspace in parameter space where new critical points emerge; a new potential well may be formed when crossing this separatrix, thus producing another stability location for excited states to settle. The \textit{Maxwell set} is the subspace in parameter space where two or more minima (maxima) have the same value; this corresponds to a point of bistability and when crossing this separatrix the global minimum changes from one well to the other. According to Ehrenfest's classification this is a first order quantum phase transition as the semiclassical potential has a discontinuity in the first derivative of the potential as a function of the varying parameter.

Calculating the expectation value of the Hamiltonian (\ref{1.1}) in the coherent state basis (\ref{1.3}) in both angular and cartesian coordinates we obtain:
\begin{widetext}
\begin{eqnarray}\label{1.4}
V(\theta,\phi;D,E,B_i) &=& \frac{D}{4}S(2S-1)\cos 2\theta + \frac{E}{2}S(2S-1)\cos 2\phi\sin^2 \theta \cr
& & + g\mu_B S \left(-B_z \cos \theta + B_x\cos\phi \sin\theta + B_y \sin\phi \sin\theta \right) + \frac{D}{4}S(2S+1) \cr
& & +\frac{1}{8}S(2S-1)(2S-2)(2S-3)\left(\frac{B_4^0}{8}(35\cos 4\theta +20\cos 2\theta +9) \right. \cr
& &\left.  + \frac{B_4^2}{2}(7\cos 2\theta +5)\cos 2\phi \sin^2\theta - B_4^3 \cos 3\phi \cos\theta \sin^3\theta + B_4^4 \cos 4\phi \sin^4\theta \right),
\end{eqnarray}
and in the variable $x$ and $y$ it is
\begin{eqnarray}\label{1.5}
V(x,y;D,E,B_i) &=&- \frac{D}{2}S(2S-1)(x^2+y^2) + \frac{E}{2}S(2S-1)(x^2-y^2) \cr
& & + g\mu_B S \left(\pm B_z \sqrt{1-x^2-y^2} + B_x  x + B_y y \right) \cr
& & +\frac{1}{8}S(2S-1)(2S-2)(2S-3)\left(B_4^0(35(x^4+y^4)-40(x^2+y^2) +70x^2y^2+8) \right. \cr
& &\left. - B_4^2 (x^2-y^2)(7(x^2+y^2)-6) \mp B_4^3x(x^2-3y^2)\sqrt{1-x^2-y^2} \right. \cr
& & \left. + B_4^4(x^4+y^4-6x^2y^2) \right)
+ D S^2.
\end{eqnarray}
\end{widetext}

As a start, in order to apply the methods of catastrophe theory directly, we will consider that the $y$ component of the external magnetic field is zero, i.e. $B_y=0$, so that the magnetic field vector is confined in the $xz$-plane, with non-zero components parallel to the easy and hard axes. The introduction of the $B_y$ component makes the application of the catastrophe theory not as straightforward and it will be put aside for the moment. However, worthwhile results can still be obtained in this case. With this consideration the semiclassical potentials in (\ref{1.4}) and (\ref{1.5}) have constant critical points in the $\phi$ variable: $\phi_c=0$ and $\pi$, which correspond to the value $y_c=0$. Then the potentials in (\ref{1.4}) and (\ref{1.5}) can be treated as one-dimensional functions for the study of their critical points. From now on we focus on the potential in angular coordinates, noting that both lead to similar results, and obtain:
\begin{widetext}
\begin{eqnarray}\label{1.6}
V(\theta,\phi_c;r_i) &=& \frac{1}{4}S(2S-1) r_3 \cos 2\theta  + g S \left(-r_2 \cos \theta \pm r_1 \sin\theta  \right) \cr
& & + \frac{1}{64}S(2S-1)(2S-2)(2S-3)\left(r_4 \cos 4\theta  \mp r_5 (2\sin 2\theta - \sin 4\theta) \right) \cr
& & + \frac{D}{4}S(2S+1) + \frac{E}{4}S(2S-1)  + \frac{1}{64}S(2S-1)(2S-2)(2S-3)(9B_4^0 + 3B_4^2 + 3B_4^4 ), 
\end{eqnarray}
\end{widetext}
where we defined the new parameters $r_i$ as:
\begin{eqnarray}\label{1.7}
r_1 &=& \mu_BB_x \cr
r_2 &=& \mu_BB_z \cr
r_3 &=& D-E+\frac{1}{16}(2S-2)(2S-3)(20B_4^0+4B_4^2-4B_4^4) \cr
r_4 &=& 35 B_4^0 -7B_4^2 + B_4^4 \cr
r_5 &=& B_4^3 .
\end{eqnarray}
The initial problem of dealing with a Hamiltonian dependent on $9$ parameters reduces the study of the critical points of a one-dimensional potential dependent on $5$ parameters. The methods of catastrophe theory are used in the semiclassical potential defined in (\ref{1.6}) to construct separatrices in parameter space $(r_1,r_2,r_3,r_4,r_5)$. By direct application of the results in Appendix A in Ref. \cite{lohrrobles2021a} we can construct the bifurcation set and Maxwell set separatrices in regions of interest.

Then, in practice, the value of the anisotropy parameters $(D,E,B_4^k)$ for a specific SMM puts us in a particular place in parameter space, while the varying strength of the external magnetic field produce trajectories therein that crosses some of the separatrices and thus the structure of the energy levels is determined accordingly.

\section{Application of the results to some examples of SMMs}\label{section3}
In order to illustrate the application of the results presented in the previous section, we focus on applying the methods in SMM complexes with a rich abundance of research and available data. For this purpose we choose three different SMMs: $\mathrm{Fe_4}$, $\mathrm{Fe_8}$, and $\mathrm{Mn_{12}}$.

In this section we will construct the parameter spaces for the chosen examples and investigate the effects of increasing the strength of the magnetic field parallel to the easy axis while having a transverse magnetic field of different magnitudes.

\subsection{$\mathrm{Fe_4}$ complexes}
The $\mathrm{Fe_4}$ SMMs are candidates for use in quantum technologies given its facility to construct different SMMs using various ligands, which changes the symmetry of the molecules and gives different values of the anisotropy parameters \cite{accorsi2006a,gregoli2009a,mannini2010a,elhallak2012a,burgess2015a,cini2018a}. As we will see, the trigonal symmetry of some of these SMMs introduces the $B_4^3$ parameter, which plays an important role in the results related to QPTs.

\begin{figure}[ht]
\includegraphics[width=\columnwidth]{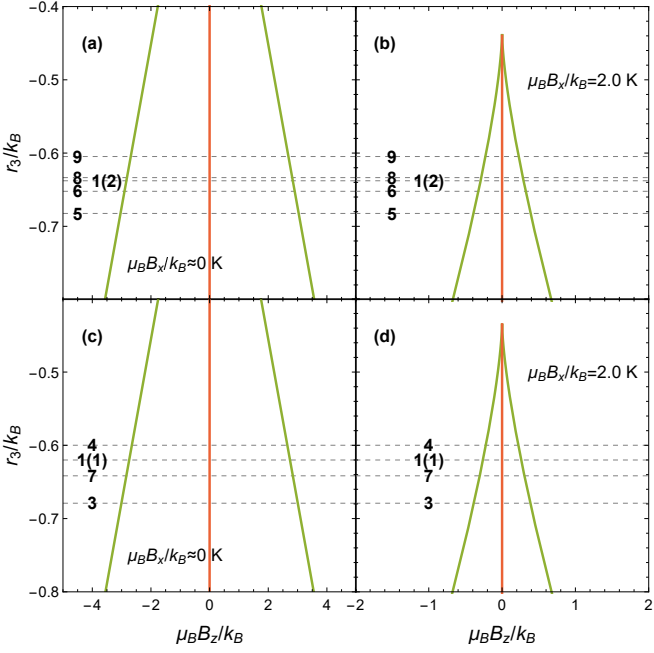}
\caption{\label{fig.1} Parameter space $(\mu_BB_z/k_B,r_3/k_B)$ for the compounds in Table \ref{table1}, depicted as horizontal dashed lines. The bifurcation set is depicted in green, and the Maxwell set in red. (a) and (c) $\mu_B B_x/k_B\approx 0$ K. (b) and (d) $\mu_B B_x/k_B =2.0$ K. (a) and (b) $r_4/k_B=0.0005$ K. (c) and (d) $r_4/k_B=0.00076$ K. Given the uncertainty of $\pm 0.5\times 10^{-5} \mathrm{cm^{-1}}$ for the $B_4^0$ parameter \cite{gregoli2009a}, we used the approximation $B_4^0=1.0\times 10^{-5} \mathrm{cm^{-1}}$ and $B_4^0=1.5\times 10^{-5} \mathrm{cm^{-1}}$, for the top and bottoms plots, respectively.}
\end{figure}

We focus on the different $\mathrm{Fe}_4$ SMMs reported in \cite{gregoli2009a}. There, they made three new tetrairon(III) $\mathrm{Fe}_4$ SMMs prepared by using different ligands, and additionally they also present the results of 8 other already studied tetrairon(III) SMMs \cite{barra1999a,cornia2004a,accorsi2006a,barra2007a,cornia2008a,condorelli2008a}. Each of the SMMs are described by distinct anisotropy parameters, and they were determined with HFEPR spectroscopy. This SMM has a ground state spin of $S=5$. In Table \ref{table1} we show the parameter values for the $12$ compounds presented in \cite{gregoli2009a}, and write here the chemical formulae of the SMMs: $[\mathrm{Fe}_4 (\mathrm{L})_2 (\mathrm{dpm})_{6}]$ with a ligand $\mathrm{H_3L}= \mathrm{R'OCH_2C}$-$(\mathrm{CH_2OH})_3$, with: ($\mathbf{1}$) $\mathrm{R'}$=allyl, ($\mathbf{2}$) $\mathrm{R'}$=($R$,$S$)-2-methyl-1-butyl, ($\mathbf{3}$) $\mathrm{R'}$=($S$)-2-methyl-1-butyl, ($\mathbf{4}$) $\mathrm{H_3L}$=11-(acetylthio)-2,2-bis(hydroxymethyl)-undecan-1-ol; with ligand a $\mathrm{H_3L}= \mathrm{RC (\mathrm{CH_2OH})_3}$ with: ($\mathbf{5}$) $\mathrm{R=Me}$, ($\mathbf{6}$) $\mathrm{R=(CH_2)_7CH=CH_2}$, ($\mathbf{7}$) $\mathrm{R=CH_2OPh}$, ($\mathbf{8}$) $\mathrm{R=Ph}$, ($\mathbf{9}$) R=4-Cl-Ph, ($\mathbf{10}$) $\mathrm{R=CH_2O(CH_2)_4(C_{16}H_9)}$; and with ligand $\mathrm{H_3L'}= t\mathrm{BuC (\mathrm{CH_2OH})_3}$ there are two with formulae: ($\mathbf{11}$) $[\mathrm{Fe}_4 (\mathrm{L'}) (\mathrm{OEt})_3 (\mathrm{dpm})_{6}]$, and ($\mathbf{12}$) $[\mathrm{Fe}_4  (\mathrm{OMe})_6 (\mathrm{dpm})_{6}]$. For the compound ($\mathbf{1}$) there are two different molecular structures reported, and they are labelled respectively as ($\mathbf{1} (1)$) and ($\mathbf{1} (2)$) in Table \ref{table1} and Fig. \ref{fig.1}. For more information about the compounds see Ref. \cite{gregoli2009a} and references therein.

\begin{table}[ht]
\caption{\label{table1}Values of the anisotropy parameters in Hamiltonian (\ref{1.1}) for $\mathrm{Fe_4}$ SMMs found in \cite{gregoli2009a}.}
 \begin{ruledtabular}
  \begin{tabular}{ c  c  c  c }
 Compound & $D/k_B [\mathrm{K}]$  & $E/k_B [\mathrm{K}]$  & $10^{5} B_4^0/k_B [\mathrm{K}]$   \\ \hline 
($\mathbf{1} (1)$) & $-0.6$ & 0.022 & 1.87\\ 
($\mathbf{1}(2)$) & $-0.626$ & 0.013 & 1.3\\
($\mathbf{2}$) & $-0.646$ & 0.043 & 3.45\\
($\mathbf{3}$) & $-0.636$ & 0.0446 & 2.3\\
($\mathbf{4}$) & $-0.593$ & 0.0086 & 2.59\\
($\mathbf{5}$) & $-0.64$ & 0.0 & 1.439\\
($\mathbf{6}$) & $-0.624$ & 0.0288 & 1.439\\
($\mathbf{7}$) & $-0.623$ & 0.02 & 2.16\\
($\mathbf{8}$) & $-0.601$ & 0.033 & 1.15\\
($\mathbf{9}$) & $-0.591$ & 0.0143& 1.58\\
($\mathbf{10}$) & $-0.588$ & 0.0115 & 3.45\\
($\mathbf{11}$) & $-0.388$ & 0.0 & $<0.719$\\
($\mathbf{12}$) & $-0.296$ & 0.0143 & $-1.58$\\
\end{tabular}
\end{ruledtabular}
\end{table}

\begin{figure}[ht]
\includegraphics[width=\columnwidth]{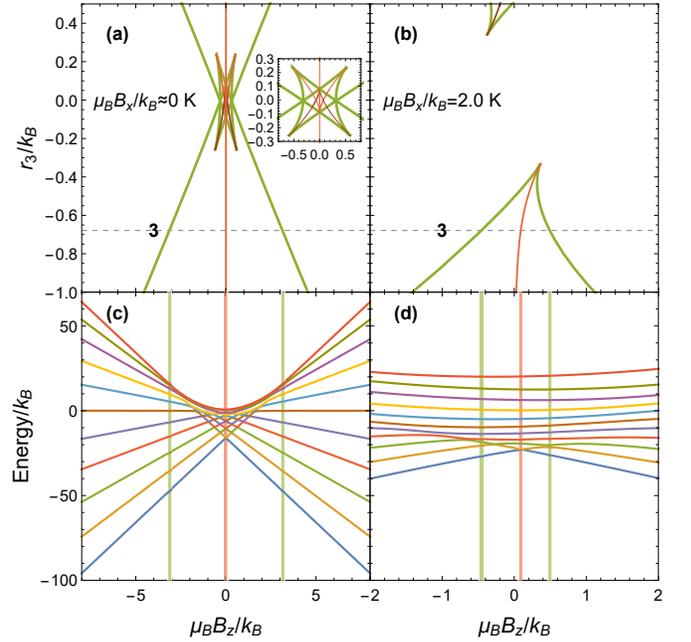}
\caption{\label{fig.2} Parameter space $(\mu_B B_z/k_B,r_3/k_B)$ for the compound ($\mathbf{3}$) in Table \ref{table1}, depicted as an horizontal line, with an additional parameter $r_5/k_B=B_4^3/k_B=0.01$ K as considered in \cite{gregoli2009a}: (a) $\mu_B B_x/k_B\approx 0$ K, and (b) $\mu_B B_x/k_B =2.0$ K. The bifurcation set is shown in green. When two minima (maxima) have the same value the Maxwell set is shown in red (darker red). The bottom plots are the corresponding eigenvalues of the Hamiltonian (\ref{1.1}): (c) $\mu_B B_x/k_B\approx 0$ K, and (d) $\mu_B B_x/k_B =2.0$ K. In vertical lines are the values of $\mu_BB_z/k_B$ for the bifurcation set (green) and Maxwell set (red). The parameters that define the system are: $S=5$, $r_3/k_B=-0.679$ K, $r_4/k_B=0.00076$ K, and $r_5/k_B=0.01$ K.}
\end{figure}

In Fig. \ref{fig.1} we show the parameter space $(\mu_BB_z/k_B,r_3/k_B)$, with $k_B$ the Boltzmann constant, when $\mu_BB_x/k_B\approx 0$ K for nine of the compounds in Table \ref{table1}, each of them depicted as an horizontal dashed line. When the strength  of the component $B_z$ is varied the bifurcation set (green) is crossed at a different value for each of the compounds, while the Maxwell set (red) is crossed at $\mu_BB_z/k_B= 0$ K for all of them. In the plots on the right, as the strength of the $B_x$ component is increased, the separatrices in parameter space change and the second potential well appears at a smaller value of $B_z$ for all cases, altering the structure of the Hamiltonian eigenvalues.

In Ref. \cite{gregoli2009a}, they propose the inclusion of the higher order anisotropy term $B_4^3$ in order to replicate the observed value of the energy barrier for compound ($\mathbf{3}$), given the trigonal symmetry of the molecule. Using the relations in (\ref{1.7}), the values of the parameters written in Table \ref{table1} for this compound, together with the value $B_4^3=0.01$ K mentioned in Ref. \cite{gregoli2009a}, map to the following values of the $r_i$ parameters of the semiclassical potential: $r_3/k_B=-0.679$ K, $r_4/k_B=0.0076$ K, and $r_5/k_B=0.01$ K. The value of the parameter $r_5=B_4^3$ is big enough to produce some significant change in the structure of the separatrices in parameter space, where the symmetry around the value of $\mu_BB_z/k_B= 0$ K is lost, as can be seen in Fig. \ref{fig.2}. As the strength of the $B_x$ component is increased the ground state first order QPT occurs at a non-zero value of $B_z$. In Ref. \cite{mannini2010a} a different $\mathrm{Fe}_4$ SMM is studied and a value of $B_4^3$ of the same order is used to recreate experimental data, which leads to a similar result as the one shown in Fig. \ref{fig.2}. When the transverse magnetic field is zero, shown in Fig. \ref{fig.2}(a), for the parameter values of the compound ($\mathbf{3}$) the bifurcation set is crossed at $\mu_BB_z/k_B=\pm 3.136$ K and the Maxwell set is crossed at $\mu_BB_z/k_B\approx 0$ K. Applying a transverse magnetic field breaks this symmetry, for example, when $\mu_B B_x/k_B=2.0$ K, as shown in Fig. \ref{fig.2}(b), the bifurcation set is crossed at $\mu_BB_z/k_B=-0.455$ K and $\mu_BB_z/k_B=0.493$ K, while the Maxwell set is crossed at the non-zero value $\mu_BB_z/k_B=0.09$ K.

These results show how a semiclassical approach and the use of catastrophe theory can be used in SMM examples to schematically describe their behavior.

\begin{figure}[ht]
\includegraphics[width=\columnwidth]{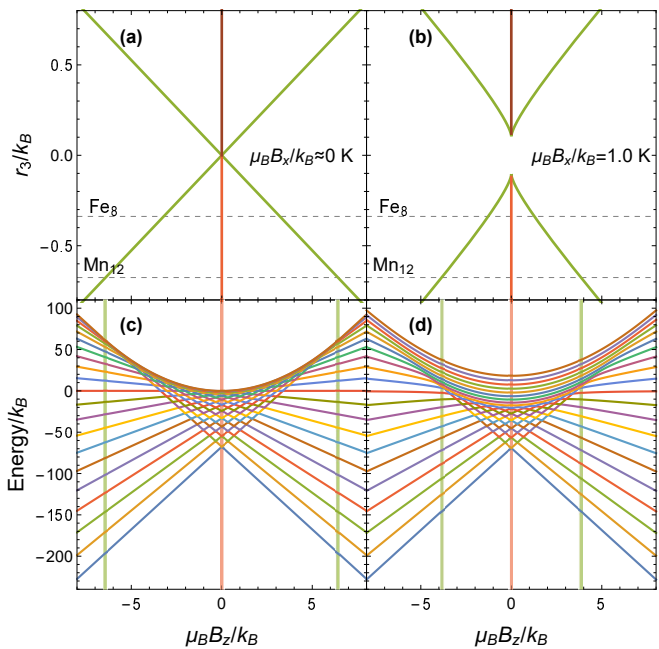}
\caption{\label{fig.3} Parameter space $(\mu_BB_z/k_B,r_3/k_B)$ for the SMMs $\mathrm{Fe_8}$ and $\mathrm{Mn_{12}}$, depicted as horizontal dashed lines, with: (a) $\mu_BB_x/k_B\approx 0$ K, and (b) $\mu_BB_x/k_B =1.0$ K. The bifurcation set is shown in green. When two minima (maxima) have the same value the Maxwell set is shown in red (darker red). The values of the parameters used for each SMM are in Table \ref{table2}, with all fourth order parameters set to zero. As an example in the bottom plots we show the eigenvalues of the Hamiltonian (\ref{1.1}) for the $\mathrm{Mn_{12}}$ SMM, for: (c) $\mu_BB_x/k_B\approx 0$ K, and (d) $\mu_BB_x/k_B =1.0$ K. In vertical lines the values of $\mu_BB_z/k_B$ corresponding to the bifurcation set (green) and Maxwell set (red) are depicted.}
\end{figure}

\subsection{$\mathrm{Fe_8}$ and $\mathrm{Mn_{12}}$ complexes}
The phenomenon of quantum tunneling of the magnetizations (QTM) has been investigated in $\mathrm{Mn_{12}}$ SMMs \cite{thomas1996a,barra1997a} and the magnetization hysteresis loops are reported to present steps at regular intervals \cite{friedman1996a}. In a $\mathrm{Mn_{12}}$-based wheel modelled as two interacting SMMs the QTM was studied in \cite{ramsey2008a}. The importance of the fourth order terms in the Hamiltonian to explain this phenomena has also been studied \cite{barra1997a,mirebeau1999a,zhong2000a,kent2000a}. By the application of pressure parallel or perpendicular to the easy axis of the SMM the properties of QTM were shown to change \cite{atkinson2017a}, which induce a change in the second and fourth order anisotropy parameters.

Studies of similar properties have been 
performed in the $\mathrm{Fe_8}$ SMMs 
\cite{barra1996a,sangregorio1997a,ohm1998a,caciuffo1998a,wernsdorfer2000a}, which is also a very important material in the development of the experimental and theoretical advancements in the subject. Different complexes of this SMM have been synthesized and compared \cite{barra2001a}. The application of microwave radiation on the $\mathrm{Fe_8}$ SMM were shown to produce changes in its magnetization 
\cite{bal2005a,petukhov2007a}. 

Many examples of SMMs are also available for testing using these methods, in particular we mention here the new results in the study of the $\mathrm{Fe_6}$ SMM in \cite{hernandez2015a,nehrkorn2021a}.

In Table \ref{table2} we list 
the parameter values of Hamiltonian (\ref{1.1}) for the following SMMs: (i) $[\mathrm{Fe}_8 \mathrm{O}_2 (\mathrm{OH})_{12}(\mathrm{tacn})_{6}]^{8+}$ in \cite{wernsdorfer2000a}, (ii) $[\mathrm{Mn}_{12} \mathrm{O}_{12} (\mathrm{CH}_3\mathrm{COO})_{16^-}(\mathrm{H}_2\mathrm{O})_4]2\mathrm{CH}_3\mathrm{COOH}\cdot 4\mathrm{H}_2\mathrm{O}$ in \cite{hill1998a}, and (iii) $[\mathrm{Bu_4N}]$-$[({}^{\mathrm{H}}\mathrm{L})_2\mathrm{Fe_6(dmf)_2]}$ in \cite{nehrkorn2021a}.

\begin{table}[ht]
\caption{\label{table2} Values of the parameters in the Hamiltonian (\ref{1.1}) for the examples of SMMs mentioned in the text.}
\begin{ruledtabular}
  \begin{tabular}{ c  c  c  c  c  c  c }
& $S$ & $D/k_B [\mathrm{K}]$ & $E/k_B [\mathrm{K}]$ & $B_4^0/k_B [\mathrm{K}]$ &  $B_4^4/k_B [\mathrm{K}]$ & Ref. \\ \hline 
(i) & 10 & $-0.292$ & $0.046$ & - & $-2.9 \times 10^{-5}$ & \cite{wernsdorfer2000a}\\ 
(ii) & 10 & $-0.676$ & - & $0.251 \times 10^{-4}$ & $-1.18 \times 10^{-4}$ & \cite{hill1998a}\\
(iii) & 19/2 & $-0.73$ & $0.129$ & $3.31 \times 10^{-4}$ & - & \cite{nehrkorn2021a}\\  
  \end{tabular}
  \end{ruledtabular}
\end{table}

For illustrative purposes the parameter space $(\mu_BB_z/k_B,r_3/k_B)$ for examples (i) and (ii) in Table \ref{table2}, which are the most representative examples of SMMs, are shown in Fig. \ref{fig.3}. We can see how the crossing of the bifurcation set and Maxwell set dictates the behavior of the energy levels of the Hamiltonian.

\begin{figure}[ht]
\includegraphics[width=\columnwidth]{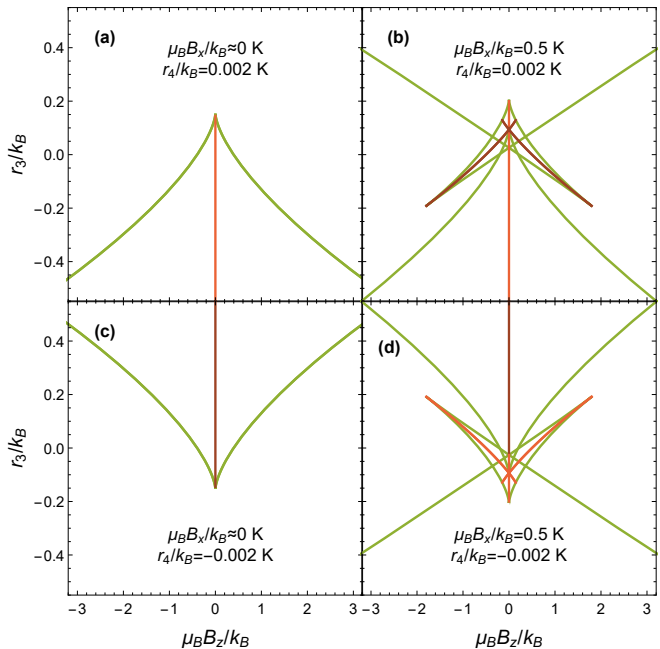}
\caption{\label{fig.4} Parameter space $(\mu_BB_z/k_B,r_3/k_B)$ for a possible example of a SMM with tetragonal symmetry and with ground state $S=10$. (a) and (c) $\mu_BB_x/k_B\approx 0$ K. (b) and (d) $\mu_BB_x/k_B =0.5$ K. (a) and (b) $r_4/k_B=0.002$ K. (c) and (d) $r_4/k_B=-0.002$ K. The bifurcation set is shown in green. When two minima (maxima) have the same value the Maxwell set is shown in red (darker red).}
\end{figure}

\subsection{Effect of fourth order parameters}\label{section4}
In the previous section we could see the effect of the fourth order parameters in the parameter space separatrices for one of the examples of a SMM with trigonal symmetry. The Maxwell set is crossed at a non-zero value for the $z$ component of the magnetic field. In this subsection we wish to perform an schematic study of the possibilities for the parameter spaces separatrices when the fourth order parameters have a more relevant contribution for the different symmetries already discussed. This will provide information of whether similar phenomena are possible for the other symmetries, and if so, for what values of the anisotropy parameters is this achieved.

For the cases of a molecule with tetragonal and orthorhombic symmetries the non-zero fourth order parameters $B_4^k$ are: $k=0,4$ and $k=0,2,4$, respectively. Both of these cases contribute to the semiclassical potential parameters $r_3$ and $r_4$, with $r_5=0$. In Fig. \ref{fig.4} we show various examples in parameter space $(\mu_BB_z/k_B,r_3/k_B)$ for this case. In the top plots we set $r_4=35 B_4^0 -7B_4^2 + B_4^4 >0$, and in the bottom plots $r_4<0$. The plots in the left and in the right show the effect of an applied transverse magnetic field. When $r_4<0$, and in the presence of a transverse magnetic field, there is a Maxwell set for the global minimum of the potential for non-zero values of $B_z$, as can be seen in Fig. \ref{fig.4}(d), which indicates the possibility of a first order QPT for the appropriate values of the parameters.

For the case of a molecule with trigonal symmetry the non-zero fourth order parameters $B_4^k$ are: $k=0,3$. An example of this case can be seen in Fig. \ref{fig.2}, where at zero transverse magnetic field, an interesting structure can be seen for small values of $r_3$ near the origin, in the inset of Fig. \ref{fig.2}(a), but far from the values of the trigonal compound (${\bd 3}$). As the transverse magnetic field is increased, this structure in parameter space disappears, and the familiar cusp appears, but now it is not centered around $\mu_BB_z/k_B=0$ K, as can be seen in Fig. \ref{fig.2}(b).

\begin{figure*}[ht]
\includegraphics{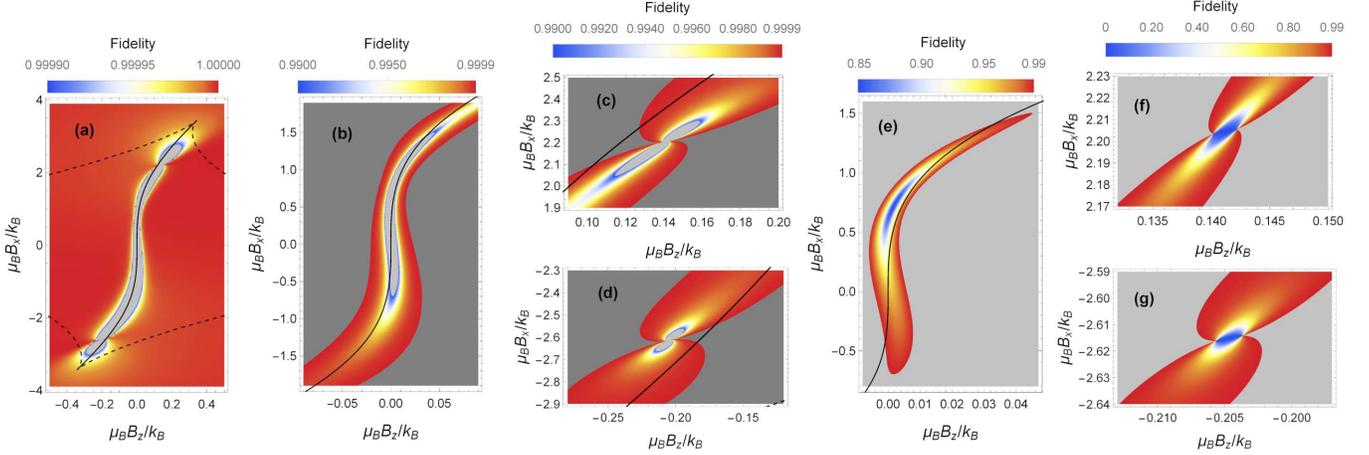}
\caption{\label{fig.5} Fidelity density plots of the ground state in the $(\mu_BB_z/k_B,\mu_BB_x/k_B)$ parameter space. The Maxwell set (black, solid) and bifurcation set (black, dashed) are also shown. Here we used $d B_z = dB_x=0.001$. (a) Full view of the fidelity in the vicinity of the Maxwell set. In the light gray area the fidelity is less than 0.9999. (b), (c), and (d) In the dark gray area the fidelity is more than 0.9999, while in the light gray area it is less than 0.99. (e), (f), and (g) Zoom to the regions of pronounced fidelity drop; in the light gray area the fidelity is more than 0.99. The parameters that define the system are: $S=5$, $r_3/k_B=-0.679$ K, $r_4/k_B=0.00076$ K, and $r_5/k_B=0.01$ K.}
\end{figure*}

\section{Fidelity}\label{section5}
The crossing of the Maxwell set, as the external magnetic field is varied, produces a sudden change in the ground state energy of the system. This is characterized as a first order QPT. In this section we are going to choose the compound $({\bd 3})$ in Table \ref{table1} of a $\mathrm{Fe_4}$ SMM with parameter: $D/k_B=-0.636$ K, $E/k_B=0.0446$ K, $B_4^0/k_B=2.3\times 10^{-5}$ K, and additionally consider $B_4^3/k_B=0.01$ K as mentioned in the previous section, as an example to study the QPTs found in parameter space as a function of the $x$ and $z$ components of the external magnetic field. To this end we will consider the fidelity of the ground state. 

Initially a concept of information theory, the fidelity has been used successfully to detect QPTs in quantum many-body systems \cite{zanardi2006a,buonsante2007a,you2007a,tzeng2008a,gu2010a,tian2011a,plotz2011a,rams2011a}. The ground state fidelity is defined as the square of the absolute value of the overlap between two ground states wave functions that differ by a small parameter:
\begin{equation}\label{5.1}
F(\lambda,d\lambda) = |\langle \psi_0 (\lambda - d \lambda)|\psi_0 (\lambda + d \lambda) \rangle|^2 ,
\end{equation}
where $\lambda$ is the parameter related to the QPT and $d\lambda$ is a small increment \cite{gu2010a}. In our case $\lambda$ are the components of the external magnetic field.

At the critical point of the QPT there is an avoided level crossing of the ground state with the first excited level, giving rise to a mixing of states, and the ground state eigenfunction changes drastically. This is seen as a sudden drop in the fidelity. 

In Fig. \ref{fig.5} we show the fidelity of the ground state of the chosen example as a density plot in $(\mu_BB_z/k_B,\mu_BB_x/k_B)$ parameter space, with an increment of $d B_z = dB_x=0.001$. The Maxwell set and the bifurcation set are depicted in black as a solid line and a dashed line, respectively. It can be seen in Fig. \ref{fig.5}(a) how the Maxwell set provides a very useful guide to signal the drop of the fidelity. In most of the region the drop of fidelity is small, but there are three smaller regions that show a very pronounced drop as can be seen in Figs. \ref{fig.5}(b), \ref{fig.5}(c), and \ref{fig.5}(d). When $\mu_BB_z/k_B\approx 0$ K and in the range of $\mu_BB_x/k_B$ from about $0.3$ K to about $1.0$ K, where the Maxwell set starts to curve, shown in Fig. \ref{fig.5}(e), the fidelity drops from 1 to 0.85. The other regions are in the neighbourhood of the points $(\mu_BB_z/k_B,\mu_BB_x/k_B)=(0.141\;\mathrm{K},2.205\;\mathrm{K})$ and $(\mu_BB_z/k_B,\mu_BB_x/k_B)=(-0.205\;\mathrm{K},-2.615\;\mathrm{K})$, where the fidelity drops from 1 to 0, as shown in Figs. \ref{fig.5}(f) and \ref{fig.5}(g), respectively. These two really pronounced drops, which are near the cusps of the separatrices, can be explained as the values of the parameters where the energy gap between the ground state and the first excited state is the lowest. Another interesting thing about the plots of the fidelity in Fig. \ref{fig.5} is that the symmetry present in the Maxwell set is not exhibited in the fidelity for the positive and negative values of the magnetic field components. This can be explained by the presence of the constant terms accompanying the $B_4^0$ parameter in (\ref{1.2}) of the fourth order operator in the Hamiltonian, which influence the ground state energy of the system and, consequently, the size of the energy gap with the first excited state.

\begin{figure*}[ht]
\includegraphics{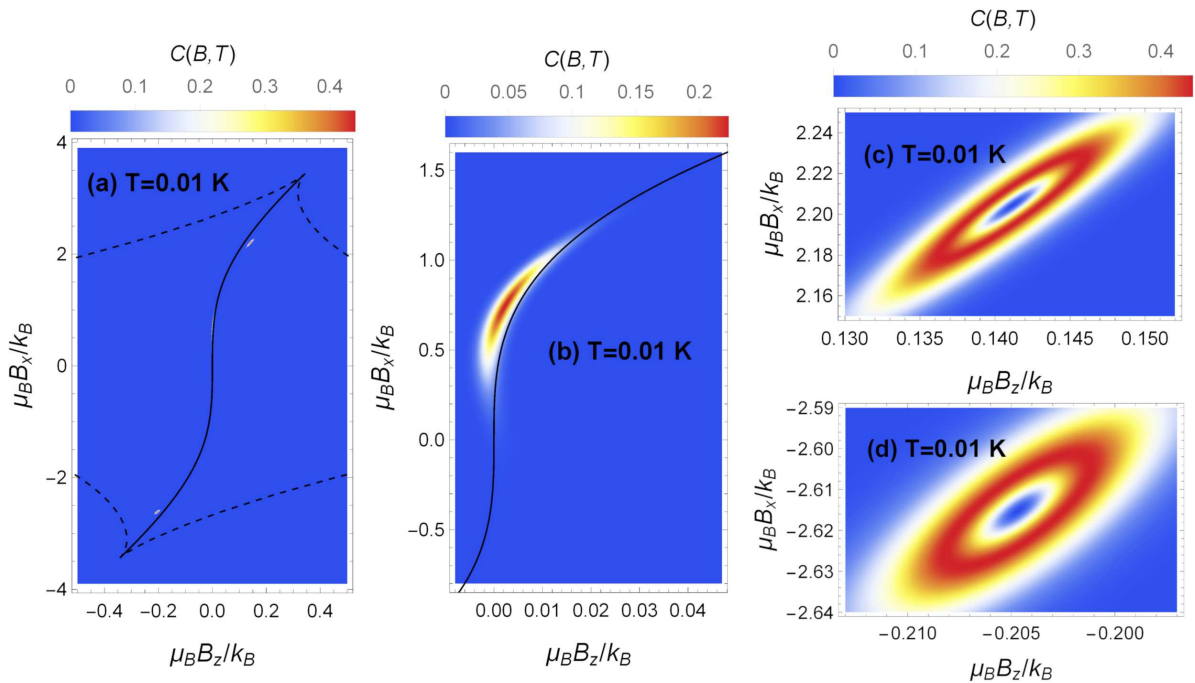}
\caption{\label{fig.6} Heat capacity density plots for the temperature value $T=0.01$ K in the $(\mu_B B_z/k_B,\mu_BB_x/k_B)$ parameter space. The Maxwell set (black, solid) and bifurcation set (black, dashed) are also shown. (a) Full view of the separatrices. (b) Zoom to a region of high heat capacity near the Maxwell set. (c) and (d) Ring-shaped regions of high heat capacity near the cusps of the separatrices. The parameters that define the system are: $S=5$, $r_3/k_B=-0.679$ K, $r_4/k_B=0.00076$ K, and $r_5/k_B=0.01$ K.}
\end{figure*}

\section{Finite temperature phase transitions}\label{section6}
Signals of quantum phase transitions can also be search for in thermodynamic quantities. At low temperature the change in the ground state energy should present itself as a sudden change in an appropriate quantity for the value of the parameters in the vicinity of the Maxwell set.

The heat capacity in $\mathrm{Fe_8}$ and $\mathrm{Mn_{12}}$ SMMs in the presence of a varying magnetic field has been measured and studied in \cite{fominaya1997a,fominaya1999a,miyazaki2001a}. Peaks in the heat capacity as a function of the magnetic field were detected at various temperatures. The localization of the peaks correspond to the value of magnetic field were the QTM takes place. In Refs. \cite{luis2000a,luis2001a,mettes2001a} they performed an experimental study of the $\mathrm{Fe_8}$ and $\mathrm{Mn_{12}}$ SMMs when a transverse magnetic field to the easy magnetization axis is applied, and measure the specific heat of the compounds at low temperature. A review of calorimetric studies in molecular magnets can be found in Ref. \cite{sorai2013a}. In Refs. \cite{fukuoka2016a,fukuoka2019a} they study the heat capacity of a molecular magnetic superconductor, where they applied a magnetic field parallel to the easy magnetization axis of the molecule, as well as perpendicular to it. In a future study it might be interesting to extend the present results to similar physical systems described by spin Hamiltonians.

The canonical partition function is defined as:
\begin{equation}\label{3.1}
Z(T) =\sum_{n}e^{-\frac{E_n}{k_B T}},
\end{equation}
where $k_B$ is the Boltzmann constant and $E_n$ are the eigenvalues of the system. From this the free Helmholtz energy is defined as:
\begin{equation}\label{3.2}
F=-k_B T \ln Z(T),
\end{equation}
and the entropy is given by
\begin{equation}\label{3.3}
S=- \frac{\partial F}{\partial T} .
\end{equation}

The heat capacity is given by:
\begin{equation}\label{3.4}
C = \frac{\partial U}{\partial T} = \frac{\partial}{\partial T} (F + TS) = T \frac{\partial S}{\partial T},
\end{equation}
where for the internal energy $U$ we have the relation to the
free energy given by $U=F+TS$. 

In this section we will continue the study of the compound ($\mathbf{3}$) $\mathrm{Fe_4}$ SMM with parameters: $D/k_B=-0.636$ K, $E/k_B=0.0446$ K, $B_4^0/k_B=2.3\times 10^{-5}$ K, and $B_4^3/k_B=0.01$ K. Using these parameters the Hamiltonian (\ref{1.1}) for this compound is defined, and its eigenvalues are obtained by diagonalization. Then, we are able to obtain the partition function in (\ref{3.1}), and with it the free Helmholtz energy, entropy and heat capacity are also directly obtained. 

\begin{figure*}[ht]
\includegraphics{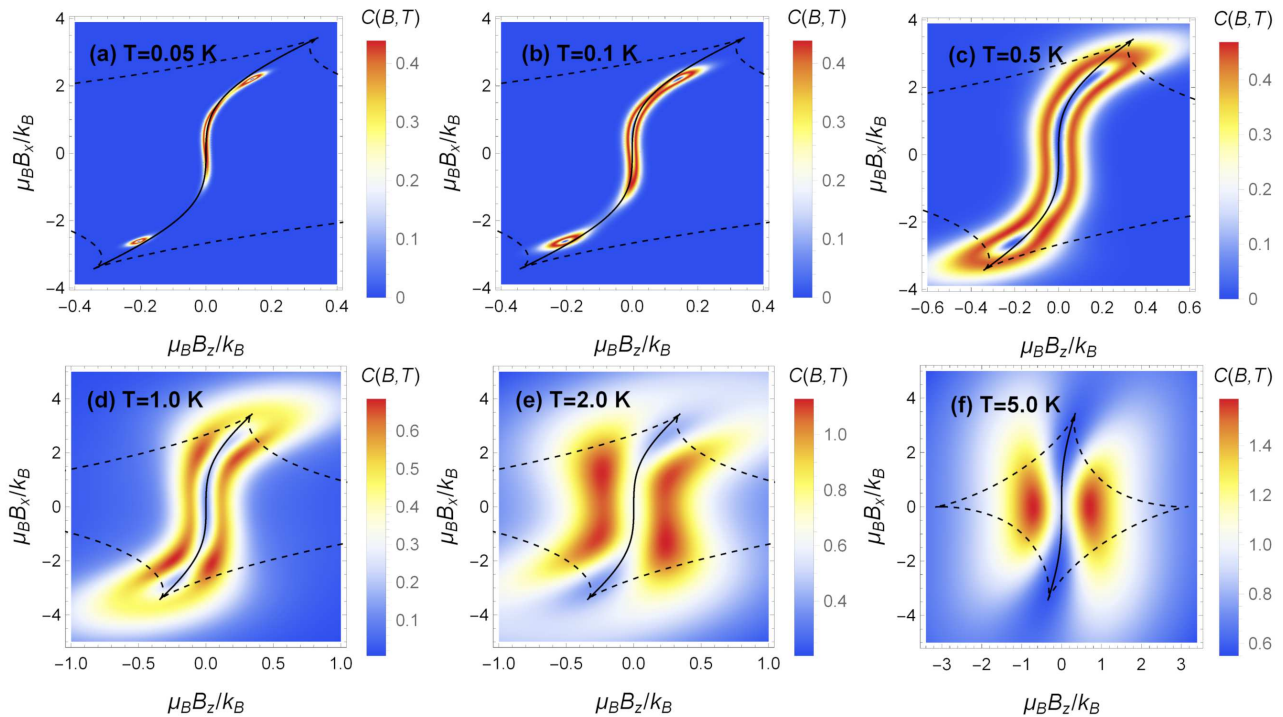}
\caption{\label{fig.7} Heat capacity density plots for different values of temperature in the $(\mu_B B_z/k_B,\mu_BB_x/k_B)$ parameter space: (a) $T=0.05$ K, (b) $T=0.1$ K, (c) $T=0.5$ K, (d) $T=1.0$ K, (e) $T=2.0$ K, and (f) $T=5.0$ K. The Maxwell set (black, solid) and bifurcation set (black, dashed) are shown. The parameters that define the system are: $S=5$, $r_3/k_B=-0.679$ K, $r_4/k_B=0.00076$ K, and $r_5/k_B=0.01$ K.}
\end{figure*}

Following the results obtained in the previous section with the fidelity, we now perform a similar analysis using the heat capacity for various values of the temperature. In Figs. \ref{fig.6} and \ref{fig.7} we show density plots of the heat capacity in the plane $(\mu_B B_z/k_B,\mu_BB_x/k_B)$ for various values of temperature. For the low temperature value $T=0.01$ K, there are peaks in the heat capacity in the vicinity of the Maxwell set shown in Fig. \ref{fig.6}(a). There are three regions of high heat capacity seen in Figs. \ref{fig.6}(b), \ref{fig.6}(c), and \ref{fig.6}(d), that coincide with the regions where there is sudden drop in fidelity. These three regions are connected following the Maxwell set, however the peaks of the heat capacity for this low temperature value are considerably smaller, but non-zero. By increasing the temperature to $T=0.05$ K, we are able to see in Fig. \ref{fig.7}(a) an increment in the heat capacity along the Maxwell set. In the same regions where the fidelity drops to zero, there is a ring-shaped area of high heat capacity, while in the interior and exterior the heat capacity is very small and goes to zero, as shown in Figs. \ref{fig.6}(c) and \ref{fig.6}(d). Then, for a fixed value of the transverse magnetic field near these two regions, the heat capacity exhibits a double peak as a function of the $z$ component of the magnetic field. As the temperature increases, the regions of high heat capacity are joined together in the shape of the Maxwell set, and now three regions can be identified: An inside region where the value of the heat capacity is small, a middle region where there are peaks of high heat capacity, and an outside region where the heat capacity is very small and goes to zero, as seen in Figs. \ref{fig.7}(b), \ref{fig.7}(c), and \ref{fig.7}(d), for temperatures $T=0.1$ K, 0.5 K, and 1.0 K, respectively. For the temperature values $T=2.0$ K and $T=5.0$ K the connected region of high heat capacity fades away and gathers into two regions on both sides of the Maxwell set and for small values of $\mu_BB_x/k_B$, as is shown in Figs. \ref{fig.7}(e) and \ref{fig.7}(f).

\section{Conclusions}\label{conclusions}
We have presented a semiclassical analysis of SMMs, described with a Hamiltonian with terms up to fourth order, utilizing the methods of catastrophe theory. 

A brief categorization of the effects of the fourth order parameter in the structure of parameter space was done, which could be useful in the design of new molecules with the appropriate values of anisotropy parameters that fall in the desired region. 

The results show the versatility of catastrophe theory for studying a system described by a Hamiltonian operator dependent on parameters and its quantum phase transitions. Separatrices in the parameter space for different SMMs with different symmetries were obtained, and it was shown how the presence of a transverse magnetic field produce changes in these separatrices. In the parameter space of the parallel and perpendicular components of the external magnetic field to the easy axis of magnetization of the molecule, we were able to construct the Maxwell and bifurcation sets for a specific $\mathrm{Fe}_4$ SMM with trigonal symmetry, which provides us with a specific set of anisotropy parameter values. The Maxwell set serves as a guide to signal a first order QPT in the ground state of the system. The fidelity was used to identify the regions of QPTs in the parameter space and they were found in the vicinity of the Maxwell set. Three interesting regions of a very substantial drop in the fidelity were detected.

It was also possible to find signals of a quantum phase transition in the heat capacity of the system at low temperatures in the same parameter space, and the Maxwell set was also important in serving as a guide for these signals. For the low temperature value $T=0.01$ K a ring-shaped region of high heat capacity was found near the cusps of the separatrices. We were also able to describe the change of the heat capacity as the temperature was increased.

We have demonstrated that catastrophe theory is extremely 
useful to describe the phase structure of a quantum system and to simplify its identification, it also provides us with the
underlying reasons for a phase 
transition and its order. Thus, it is of great importance for a better 
insight into phase transitions and their origin, and we hope that
it can lead to further predictions. 
We also hope that the results presented here are motivating for future experiments in detecting QPTs in SMMs, as well as for the application of the results to other physical systems described by similar spin Hamiltonians.

\begin{acknowledgments}
We acknowledge financial support from PAPIIT-DGAPA IN117923, IN100421.
\end{acknowledgments}

\end{document}